\documentclass[%
 arxiv
 amsmath,amssymb,
 reprint,%
]{revtex4-1}
\usepackage[english]{babel}
\usepackage{graphicx}% Include figure files
\usepackage{dcolumn}% Align table columns on decimal point
\usepackage{bm}% bold math
\usepackage[utf8]{inputenc}
\usepackage[T1]{fontenc}
\usepackage{mathptmx}
\usepackage{etoolbox}

\usepackage[version=4]{mhchem}
\usepackage{xcolor}

\usepackage{mathtools}

\makeatletter
\def\@email#1#2{%
 \endgroup
 \patchcmd{\titleblock@produce}
  {\frontmatter@RRAPformat}
  {\frontmatter@RRAPformat{\produce@RRAP{*#1\href{mailto:#2}{#2}}}\frontmatter@RRAPformat}
  {}{}
}%
\makeatother
\begin{document}

\preprint{AIP/123-QED}

\title[Detonation modeling with the Particles on Demand method]{Detonation modeling with the Particles on Demand method}
% Force line breaks with \\
\author{N. Sawant}
\author{B. Dorschner}%
% \affiliation{ 
% %\\This line break forced with \textbackslash\textbackslash
% }%

\author{I. V. Karlin}
 \email{ikarlin@ethz.ch}
 %\homepage{http://www.Second.institution.edu/~Charlie.Author.}
\affiliation{%
Department of Mechanical and Process Engineering, ETH Zurich, Zurich, 8092, Switzerland%\\This line break forced% with \\
}%

\date{\today}% It is always \today, today,

\begin{abstract}
A kinetic model based on the Particles on Demand method is introduced for gas phase detonation hydrodynamics in conjunction with the Lee--Tarver reaction model. The proposed model is realized on two- and three-dimensional lattices and is validated with a set of benchmarks. Quantitative validation is performed with the Chapman--Jouguet theory up to a detonation wave speed of Mach 20 in one dimension. Two-dimensional outward expanding circular detonation is tested for isotropy of the model as well as for the asymptotic detonation wave speed. Mach reflection angles are verified in setups consisting of interacting strong bow shocks emanating from detonation. Spherical detonation is computed to show viability of the proposed model for three dimensional simulations.
\end{abstract}

\maketitle

% \begin{quotation}
% The ``lead paragraph'' is encapsulated with the \LaTeX\ 
% \verb+quotation+ environment and is formatted as a single paragraph before the first section heading. 
% (The \verb+quotation+ environment reverts to its usual meaning after the first sectioning command.) 
% Note that numbered references are allowed in the lead paragraph.
% %
% The lead paragraph will only be found in an article being prepared for the journal \textit{Chaos}.
% \end{quotation}

\section{\label{sec:introduction}Introduction}
Detonation involves a supersonic wave which is sustained by combustion. A typical detonation structure is a shock wave followed by a deflagration, wherein the shock wave heats the reactants to a temperature at which they can react at a rate high enough for the ensuing deflagration to propagate as fast as the shock \cite{williams_combustion_1985}. Studies of detonation were initiated by Berthelot and Vieille \cite{berthelot_sur_1882} and by Mallard and Le Chatelier \cite{mallard_sur_1881}. Mathematical predictions for the propagation of detonation waves were
given by Chapman \cite{chapman_rate_1899} and Jouguet \cite{jouguet_mecanique_1917}, followed by studies on the wave structure by Zel’dovich \cite{zeldovich_16_1992}, von Neumann \cite{von_neuman_theory_1942}, and Döring \cite{grossman_contributions_1949}. 
More recently, numerical simulations based on the reactive Navier--Stokes equations were used to study detonation phenomena. 
An example of the reactive Navier--Stokes approach is the deflagration-to-detonation transition simulation of an acetylene–air mixture \cite{khokhlov_numerical_1999} with the effects of viscosity, thermal conduction and molecular diffusion accounted for in the model. Reaction is modeled by first-order chemical kinetics with constant Lewis, Prandtl and Schmidt number. A single step chemistry representing hydrogen-Air mixture combined with Navier Stokes equations have also been used for three dimensional detonation simulations \cite{gamezo_numerical_2007}. The same combination of models calibrated for ethylene-oxygen mixture and supplemented with species mass diffusivity was used to study deflagration-detonation transition with varying blockage ratios in three dimensions. In some works, simpler models have been used for hydrodynamics with more accurate chemical models. For instance, unsteady reactive Euler equations in combination with sophisticated chemical models involving NASA polynomials have been used for numerical simulations of transmitted detonation and re-ignition  in a \ce{H2}-\ce{O2}-\ce{Ar} mixture \cite{jones_numerical_2000}.  Recent axisymmetric Reynolds averaged Navier Stokes simulations with $7-$ species, $8-$ step hydrogen-air chemistry \cite{coates_numerical_2019} and three dimensional time accurate enthanol-air detonation simulation \cite{desai_direct_2021} with $28-$ species \cite{bhagatwala_direct_2014} chemistry are noteworthy developments in detailed chemistry detonation simulations. An overview of the progress in gaseous detonation experiments, modeling, and simulation has been made in [\onlinecite{shepherd_detonation_2009}] with special attention to the challenges in detonation simulation. It can be inferred from the available literature that the choice of the hydrodynamic model and of the reaction model depends on the application, the desired accuracy of the solution and practical trade-offs with respect to modelling complexity and compute cost. On the other hand, a variety of methods have been developed over the years for modelling hydrodynamics that do not make use of the traditional continuum based Navier--Stokes or Euler equations. We select one such non-traditional hydrodynamic model and evaluate its viability for detonation simulations. 
 
 A class of models for hydrodynamics  rely on the fact that the Boltzmann transport equation is equivalent to the Navier--Stokes equations \cite{chapman_mathematical_1970} when represented by a low order asymptotic expansion in powers of the Knudsen number. These mesoscopic kinetic methods solve a discrete representation of the Boltzmann transport equation with model particles and simple collision operators in order to simulate the Navier Stokes equations at the macroscopic scale. Popular examples of kinetic methods \cite{dimarco_numerical_2014} include the lattice Boltzmann method (LBM), discrete velocity method (DVM), gas kinetic scheme (GKS), Direct Simulation Monte Carlo (DSMC) among others. The LBM \cite{higuera_boltzmann_1989,higuera_simulating_1989} models fluid flow using a fully discrete kinetic system of designer particles discrete velocities $\bm{c}_i, i=0,\cdots Q-1$, fitting into a regular space-filling lattice. In the LBM, the kinetic evolution equation for the probability distribution functions $f_i(\bm{x},t)$ follows a simple algorithm of ``stream along links $\bm{c}_i$ and collide at the nodes $\bm{x}$ in discrete time $t$". The LBM has evolved into a versatile tool for the simulation of complex flows \cite{kruger_lattice_2017,sharma_current_2020,succi_lattice_2018}. 
 
 The LBM solves the Boltzmann transport equation by discretizing the velocity space and time. In LBM, the propagation of particles along the characteristic discrete velocities causes the particles to hop amongst the lattice node sites, thereby leading to an exact spatial discretization. 
 On the other hand, the DVM \cite{watari_possibility_2004,wagner_investigation_2006} discretizes the velocity space, time as well as the spatial fluxes. 
 In the DVM, the spatial discretization is performed by traditional methods such as the finite volume method or the finite difference method (FDLBM) \cite{gan_two-dimensional_2008}. 
 Within the pool of kinetic models, the FDLBM has enjoyed considerable success at modeling detonation.
 
 One of the earliest and well-known works in detonation modelling with FDLBM \cite{yan_lattice_2013} for hydrodynamics makes use of the Lee--Tarver model \cite{lee_phenomenological_1980} for modeling the chemical reactions. A large range of detonation problems are solved with quantitative validation up to Mach 3.5, thereby demonstrating feasibility of kinetic models for detonation simulations. 
 This work was followed by an exploration of detonation problems up to Mach 2, involving non-equilibrium hydrodynamics and negative temperature coefficient of reaction \cite{zhang_kinetic_2016,zhang_one-dimensional_2019}, again with the FDLBM and the Lee--Tarver combination of methods. Recently, the FDLBM coupled with a two-step chemistry model \cite{ji_three-dimensional_2022} was used to demonstrate three dimensional detonation setups and validation in one dimension for Mach 1.74. Although the progress in kinetic hydrodynamic models for detonation has been commendable, the speed of detonation waves in certain mixtures like hydrogen-Air and acetylene-Air \cite{shepherd_detonation_2009,lee_detonation_2008} are in the range of $5$ to $10$ times the speed of sound. Accurate modeling of hydrogen-Air mixtures is important from a safety standpoint as we move towards cleaner combustion for our transportation needs. Therefore, presenting a hydrodynamic model for simulation of detonation regimes of practical interest is our primary motivation. 
 However, the LBM remains limited to moderate Mach number flows due the lack of Galilean invariance and insufficient isotropy of the velocity discretization (see, e.g. \cite{saadat_extended_2021-1}). Thus, we adopt the Particle on Demand (PonD) \cite{dorschner_particles_2018} method, which describes the evolution of population relative to a local and optimal reference frame, which has shown to overcome the low Mach barrier of LBM and enables supersonic flow simulations with ease \cite{dorschner_particles_2018,kallikounis_multiscale_2021,reyhanian_thermokinetic_2020}.
 In this work, we couple the PonD for hydrodynamics with the Lee--Tarver \cite{lee_phenomenological_1980} model for reactions in order to introduce a new hydrodynamic model for detonation simulations.   

The paper is structured as follows: In section (\ref{sec:detonation_model}), a summary of the Lee--Tarver detonation model and the form of its implementation in the current work is presented. Section (\ref{sec:hydrodynamic_model}) is an essential part of the paper in which we discuss in detail the proposed hydrodynamic model based on PonD. The model is validated in section (\ref{sec:results_and_discussion}) for one dimensional and two dimensional setups followed by a three dimensional example. The conclusion section (\ref{sec:conclusion}) is followed by Appendices that contain supplementary implementation details.
\section{\label{sec:detonation_model}Reaction model}
The Lee--Tarver model \cite{lee_phenomenological_1980} for chemical reactions is represented by a generalized energy release rate equation,
\begin{align}
    &\dot \Lambda^{\text {react}} = a (1-\Lambda)^x \eta^r + b(1-\Lambda)^x \Lambda^y P^z \nonumber \\
    & \eta=\frac{\rho_1}{\rho_0}-1,
    \label{eq:LTVoriginal}
\end{align}
where $\Lambda$ is the fraction of the explosive (reactant) that has reacted, $P$ is the pressure, $\rho_1/\rho_0$ is the ratio of post to pre-shock density of the explosive, while $a,x,r,b,y$ and $z$ are constants.

In order to model the chemical reactions in our detonation simulations, we use a simplified form of the Lee--Tarver Model \cite{yan_lattice_2013} by choosing $a=0.001$ and ignoring the other constant parameters. The resulting expression for the rate of change of the product mass fraction due to chemical reaction $\dot \Lambda^{\text {react}}$, in terms of the product mass fraction $\Lambda$ is,
\begin{align}
    &\dot \Lambda^{\text {react}} = 0.001 (1-\Lambda) + (1-\Lambda) \Lambda.
    \label{eq:LTVsimplified}
\end{align}
For a setup with an initial unburnt temperature $T_{u}$, the reaction occurs only above a threshold temperature $T_R=1.1 T_{u}$, i.e. at any given node on the lattice $\dot \Lambda^{\text {react}} = 0$ if $T < T_R$.
The transport of the product mass fraction $\Lambda$ is modeled with the convection-reaction equation:
\begin{align}
    &\partial_t \Lambda + \bm{u} \cdot \nabla \Lambda = \dot \Lambda^{\text {react}}.
    \label{eq:convectionReaction}
\end{align}
The transport equation (\ref{eq:convectionReaction}) is solved by a finite difference method using the WENO5 for spatial discretization and RK4 for time integration. The velocity $\bm{u}$ in (\ref{eq:convectionReaction}) is the local macroscopic fluid velocity which is an input from the hydrodynamic solver. In this work, we use a very simple reaction model in order to focus on the validation of the proposed hydrodynamic solver for detonation regime. In principle, kinetic models with detailed chemistry and accurate diffusion \cite{sawant_consistent_2021,sawant_lattice_2021} can be coupled with the PonD kinetic model for realistic detonation, which will be the focus of future work.
\section{\label{sec:hydrodynamic_model} Kinetic model for compressible flow}
The hydrodynamics is solved by the Particles on Demand (PonD) method \cite{dorschner_particles_2018,reyhanian_thermokinetic_2020,kallikounis_multiscale_2021,kallikounis_particles_2022}. 
The LBM models the flow in a constant global thermodynamic reference frame $\lambda(\bm{u}=0,T=T_L)$. Here, $\bm{u}=0$ means that the reference frame is at rest with respect to the lab while $T=T_L$ signifies that the reference frame is at a fixed reference temperature $T_L$. The lattice temperature $T_L$ is a known characteristic of the lattice (\ref{sec:lattices:appendix}). The LBM models the macroscopic velocity and temperature as deviations from these reference values. Large deviations from the reference values can lead to instability or errors, although flows with Mach numbers as high as $2$ \cite{saadat_extended_2021-1,saadat_extended_2021} and temperature ratios as high as $10$ \cite{sawant_lattice_2021,sawant_consistent_2021-1} are feasible using more sophisticated LB models. 
In order to circumvent the aforementioned errors due to large deviations, PonD describes the evolution of the population in a local reference frame $\lambda(\bm{u}(\bm{x}, t),T(x,t))$, where $\bm{u}(\bm{x},t)$ is the local macroscopic fluid velocity and $T(\bm{x},t)$ is the local fluid temperature at a given location in space and time. 
The particle velocities $\bm{v}_i$ at any node $\bm{x}$ at discrete time $t$ in a space-filling lattice with constant discrete velocities $\bm{c}_i$ are then defined as a function of the local reference frame $\lambda(\bm{u},T)$ as,
\begin{align}
    \bm{v}_{i}(\bm{u},T) &= \sqrt{\frac{T}{T_L}} \bm{c}_{i} + \bm{u}.
    \label{eq:pondVelocity}
\end{align} 
In the PonD method, a population $f_i(\bm{x},t)$ propagates along its corresponding particle velocity $\bm{v}_i(\bm{x},t)$ in time interval $\Delta t$. The population is then transformed to be expressed in destination reference frame followed by collision. Details of the semi-Lagrangian propagation \cite{dorschner_particles_2018} and the transformation procedure \cite{kallikounis_particles_2022} will be explained in section (\ref{sec:spatial_discretization}) on propagation.

We use the two-population PonD model \cite{kallikounis_multiscale_2021} which consists of a set of populations $f_i$ that represent the mass, momentum and translational energy and an another set of populations $g_i$ for the internal energy. Although it is possible to have an adjustable Prandtl number $\rm Pr$ with the original model \cite{kallikounis_multiscale_2021}, it is not required for the setups solved in this work. Therefore, we use a simplified version of the model with a fixed Prandtl number $\rm Pr=1$. 
% %
\subsection{\label{sec:time_discretization} Collision}
During collision, the flow density $\rho$, velocity $u$ and temperature $T$ are calculated from the conserved moments of the distribution functions $f_i$ and $g_i$ as,
\begin{align}
    &\rho=\sum_i f_i, \\
    &\rho \bm{u}=\sum_i f_i \bm{v}_{i},\\
    &2 \rho E = 2 \rho \left(C_v T + \frac{\bm{u}^2}{2}\right) = \sum_i g_i + \sum_i f_i \bm{v}^2_{i}.
\end{align}
Higher order moments of $f_i$ and $g_i$ are provided in \ref{sec:moments:appendix} for completeness.
{The collision operation leading to post collision populations $f_i (\bm x, t)^{\rm pc}$ and $g_i (\bm x, t)^{\rm pc}$ is written as, 
\begin{align}
    f_i (\bm x, t)^{\rm pc} &= f_i (\bm x,t) + 2 \beta (f_i^{\rm eq}(\bm x,t) - f_i(\bm x,t)), \\
    g_i (\bm x, t)^{\rm pc} &= g_i (\bm x,t) + 2 \beta (g_i^{\rm eq}(\bm x,t) - g_i(\bm x,t)) \nonumber \\ 
    & + 2 f_i^{\rm eq}(\bm x,t) Q^{\rm r} \dot \Lambda^{\text {react}} \Delta t,    
\end{align}}
where the equilibrium distributions function are given by
\begin{align}
    &f_i^{\rm eq}(\bm x,t) = w_i \,\rho(\bm x,t), \\
    &g_i^{\rm eq}(\bm x,t) = (2 C_v - D) T(\bm x,t) f_i^{\rm eq}(\bm x,t).  
\end{align}
Here, $C_v$ is the specific heat at constant volume and $D$ is the space dimension. The heat of reaction $Q^{\rm r}$ is an input parameter that controls the heat added due to the source term $\dot \Lambda^{\text {react}}$ provided by the reaction model. The constant weights $w_i(\bm{c}_i) \sim \mathcal{N}(0,T_L)$, which are known for a lattice \cite{chikatamarla_entropy_2006} form a normal distribution in the discrete velocity space $\bm{c}_i$, with mean $0$ and standard deviation $T_L$ (\ref{sec:lattices:appendix}) . The relaxation parameter $\beta$ incorporates the kinematic viscosity $\nu$ into the model by the relation
\begin{align}
    \beta&=\frac{T \Delta t}{2 \nu + T \Delta t }.
\end{align}
\subsection{\label{sec:spatial_discretization}Propagation}
The propagation is implemented using a semi-Lagrangian approach, where the populations at location $\bm{x}$ are given by populations at the departure points of characteristic lines $(\bm{x} - \bm{v} \Delta t)$.
Each of these populations exist in their local reference frames, distinct from one another and also from the destination frame. Therefore, a frame transformation operation is necessary to express them in the destination local reference frame $\lambda(\bm{u},T)$, which will be described following the discussion on propagation. For now, we introduce a short notation for transformation (\ref{eq:gradGeneralTransform}) of a population $f_i$ from a frame $\lambda$ to $\lambda'$ as,
\begin{align}
    f_i^{\lambda'} = \mathcal{G}_\lambda^{\lambda'} f_i^{\lambda}
    \label{eq:transformationSymbolic}
\end{align}
The process of spatio-temporal discretization in the Particle-on-Demand method occurs through the propagation of the distribution functions with their corresponding particle velocities \cite{dorschner_particles_2018}. 
For a node located at $\bm{x}$ having reference frame $\lambda(\bm{u},T)$, a population $f_i(\bm{x})$ is replaced by a transformed population that would be propagated to $\bm{x}$ due to displacement with it's particle velocity $\bm{v}_i$ in time interval $\Delta t$.
    \begin{align}
        f_i^\lambda(\bm{x},t) = \mathcal{G}_{\lambda'}^{\lambda} f_i^{\lambda'}(\bm{x}-\bm{v_i}\Delta t,t-\Delta t)
        \label{eq:propagationExact}
    \end{align}
    The propagation algorithm can be summarized as follows:
\begin{enumerate}
    \item In practice, the location $(\bm{x}-\bm{v_i}\Delta t)$ does not necessarily correspond at a node location on the Cartesian lattice. Therefore, an interpolation operation \cite{reyhanian_thermokinetic_2021,rees_3d_2014} is needed in conjunction to (\ref{eq:propagationExact}). In general, with an interpolation kernel $W(\bm{x})$ acting on $p$ nodes in the neighbourhood of the departure point $\bm{x}_D=(\bm{x}-\bm{v_i}\Delta t)$, the advection (\ref{eq:propagationExact}) is rewritten as, 
    \begin{align}
        f_i^{\lambda}(\bm{x},t) = \sum_{s=0}^{p-1} W(\bm{x}_D-\bm{x}_s) \mathcal{G}_{\lambda ^s}^{\lambda} f_i^{\lambda^s}(\bm{x}_s,t-\Delta t)
        \label{eq:propagationInterpolated}
    \end{align}
    \label{step:propagationInterpolated}
    \item Once the advection step (\ref{eq:propagationInterpolated}) is performed for all the $Q$ populations at a node for both the $f_i$ as well as the $g_i$ populations, the parameters defining the reference frame, $u$ and $T$ are recalculated from the new set of populations $f_i^1$ and $g_i^1$ with the frame update step as,
    \begin{align}
        &\bm{u}_1 = \frac{\sum_i f_i^1 \bm{v}_{i }}{\sum_i f_i^1}, \\
        &T_1 = \frac{1}{C_v} \left( \frac{\sum_i f_i^1 \bm{v}_{i}^2 + \sum_i g_i^1}{2 \rho} - \frac{\bm{u}_1^2}{2} \right)
        \label{eq:recalculateFrame}
    \end{align}
    \label{step:recalculateFrame}
    \item The post advection frame $\lambda_1(\bm{u}_1,T_1)$ is compared to the pre-advection frame $\lambda(\bm{u},T)$ for convergence within some tolerances $\epsilon_{abs}=10^{-8}$ and $\epsilon_{rel}=10^{-10}$. If ($|\bm{u}_1-\bm{u}|> \epsilon_{abs} + |\bm{u}| \epsilon_{rel}$ or $|T_1-T| > \epsilon_{abs} + |T| \epsilon_{rel}$), the propagation has not converged, the advection procedure (\ref{step:propagationInterpolated}) is repeated with respect to the new frame,
    \begin{align}
        f_i^{\lambda_1}(\bm{x},t) = \sum_{s=0}^{p-1} W(\bm{x}_D^1-\bm{x}_s^1) \mathcal{G}_{\lambda ^s}^{\lambda} f_i^{\lambda^s}(\bm{x}^1_s,t-\Delta t)
        \label{eq:propagationInterpolated1}
    \end{align}
    \item Recalculate the frame parameters with the frame update step (\ref{step:recalculateFrame}). Repeat advection and frame update  until convergence.
\end{enumerate}
After the propagation procedure is performed on all the nodes in the lattice, spatio-temporal discretization is considered complete. Let us briefly discuss the nature of the transformation process (\ref{eq:transformationSymbolic}) performed by the operator $\mathcal{G}_\lambda^{\lambda'}$ in the next subsection. 
\subsubsection{\label{sec:frame_transformation} Frame Transformation}
In a lattice of $Q$ discrete velocities $\bm{c}=(\bm{c}_0,\bm{c}_1,...,\bm{c}_{Q-1})$, at any node in frame $\lambda$ , a maximum of $Q$ linearly independent moments ${(M^{\lambda}_0,M^{\lambda}_1,...,M^{\lambda}_{Q-1})}$ can be calculated from the distribution function $\bm{f}=(f_{0},f_{1},...,f_{Q-1})$. In order to construct a probability distribution function $f_i$ that satisfies a set of moments, for e.g., up to the third order $(p+q+r) \le 3$, in the particle velocity space and in a reference frame $\lambda(\bm{u},T)$,
\begin{align}
    \sum_i f_i v_{ix}^p(u_x,T) v_{iy}^q(u_y,T) v_{iz}^r(u_z,T) &= M^{\lambda}_{x^p y^q z^r},
    \label{eq:genericMomentUptoO3}
\end{align}
we make use of the Grad's Hermite polynomial expansion \cite{grad_kinetic_1949}.
The expansion is given by the series,
\begin{align}
    &f_i = w_i \sum_{n=0}^{\infty} \frac{1}{n!} a(\bm{m};\lambda(\bm{u},T))^{(n)} H_i^{(n)}.
    \label{eq:gradGeneral}
\end{align}
The coefficients $a^{(n)}(\bm{m};\lambda(\bm{u},T))$ corresponding to the constant Hermite polynomials $H_i^{(n)}$ are calculated such that they satisfy the constraints (\ref{eq:genericMomentUptoO3}) (\ref{sec:hermiteCoefficents:appendix}) upto a certain order. The coefficients are a function of two distinct set of inputs, the vector $\bm{m}$ of moments that are the moment constraints to be satisfied by the resulting distribution $f_i$, and the frame $\lambda(\bm{u},T)$ in which $f_i$ satisfies those constraints. We exploit this distinction to perform frame transformation.
In this work, since we restrict ourselves to moments upto order three, the vector ${\bm{m}^{\lambda}=(M^{\lambda}_0,M^{\lambda}_1,...,M^{\lambda}_{d_m-1})}$ consists of $d_m=10$ unique moments in two dimensions and $d_m=20$ unique moments in three dimensions, irrespective of the size of the lattice in the velocity space $Q$. In other words, any lattice can be used as long as $Q \ge d_m, \bm{m} \in \mathcal R^{d_m}$.
For transforming the populations from a frame $\lambda$ to another frame $\lambda'$, we use the matching condition of the moments \cite{dorschner_particles_2018} on $d_m$ moments, which reads,
\begin{align}
    \bm{m}^{\lambda}(\bm{f})=\bm{m}^{\lambda'}(\bm{f})
\end{align}
The moment matching rule states that the moments are independent of the choice of frame (or guage). In order to transform populations from frame $\lambda$ to frame $\lambda'$, we calculate the moments $\bm{m}^{\lambda}(\bm{f}^\lambda)$ and then generate new populations by making use of the series expansion (\ref{eq:gradGeneral}),   
\begin{align}
    &f_i^{\lambda'} = w_i \sum_{n=0}^{\infty} \frac{1}{n!} a(\bm{m}^\lambda;\lambda'(\bm{u},T))^{(n)} H_i^{(n)}
    \label{eq:gradGeneralTransform}
\end{align}
The short notation $\mathcal{G}_\lambda^{\lambda'}$ introduced in (\ref{eq:transformationSymbolic}) represents this transformation (\ref{eq:gradGeneralTransform}) elegantly as $f_i^{\lambda'} = \mathcal{G}_\lambda^{\lambda'} f_i^{\lambda}$.
\section{\label{sec:results_and_discussion}Results and discussion}
In this section, we validate the proposed model against the  Chapman Jouguet (CJ) theory of detonation waves (\ref{sec:analytical_solution:appendix}). Specifically, for the one dimensional detonation simulations, we compare the detonation wave speed, the temperature at the CJ point and the density at the CJ point with the analytical values. Next, we analyse the detonation wave speed in a two dimensional outward expanding circular detonation setup. Critical angles of Mach stem creation are also validated in two dimensional setups consisting of opposed detonation waves. We close the results section with a qualitative simulation of spherical detonation. One dimensional and two dimensional simulations have been performed with the two dimensional $D2Q16$ lattice \cite{ansumali_minimal_2003,kallikounis_multiscale_2021} whereas the spherical detonation is computed on the three dimensional $D3Q39$ lattice \cite{frapolli_theory_2020,chikatamarla_lattices_2009}. The lattice temperature, velocities and weights are written out in \ref{sec:lattices:appendix}. For the simulations performed in this section, kinematic viscosity $\nu=10^{-3}$ and an adiabatic coefficient $\gamma=1.4$ is used unless mentioned otherwise.  
\subsection{\label{sec:1d_detonation}1D detonation}
Detonation simulations have been performed on a uniform grid of $N=5000$ points (nodes), representing a one dimensional tube of length $L$ along the $x$ axis. The unburnt initial condition is applied on most of the length of the tube $(x>0.05L)$ for pressure $P_{u}$, density $\rho_{u}$ and the product mass fraction $\Lambda_u$. The remaining small left $5\%$ of the of the domain is initialized with a fully reacted burnt product mass fraction $\Lambda_b$. In order to initiate a shock, the burnt section of the domain is initialized with pressure $P_{b}$ and density $\rho_{b}$, both of which are higher than the rest of the domain. 
\begin{align}
(\Lambda,\rho,P)=\begin{cases}
\Lambda_u=0,\rho_u=1.0,P_u=0.4,& x > 0.05 L,\\
\Lambda_b=1,\rho_b=1.7,P_b=1.0,&\text{otherwise. }
\label{eq:ic1dDet} 
\end{cases}
\end{align}
We vary the non-dimensional heat of reaction $Q^{\rm r}$ between $0.25\le Q^{\rm r}\le 120$ in order to compare the detonation wave speed and the state at the CJ point with the analytical prediction. This wide range of heat of reaction produces detonation waves that travel with speeds corresponding to Mach numbers in the range of $1.56$ to $20.33$. The ratio of temperature at the CJ point to the temperature of the unburnt fluid ranges from $T_{\rm CJ}/T_u=1.39$ for $Q^{\rm r}=0.25$ to $T_{\rm CJ}/T_u=141.16$ for $Q^{\rm r}=120$.

\begin{figure}[ht]
	\centering
	\includegraphics[keepaspectratio=true,width=0.48\textwidth]{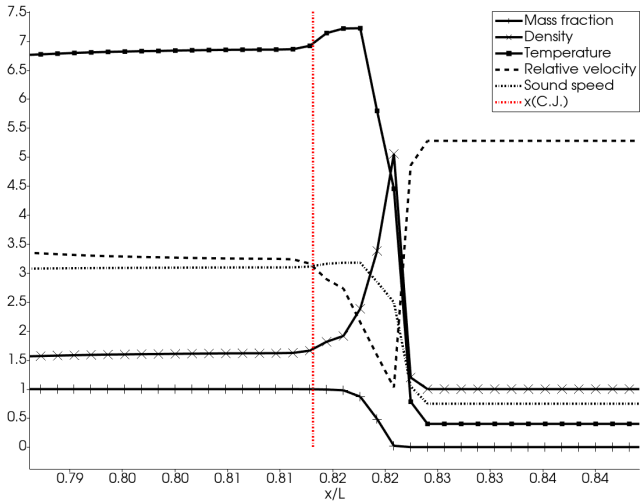}
	\caption{Profiles of density, temperature, flow velocity relative to the detonation shock speed and the local speed of sound (\ref{sec:analytical_solution:appendix}). The heat of reaction is $Q^{\rm r}=14$.} 
	\label{fig:1d_detonation_profiles}
\end{figure}

\begin{figure}[ht]
	\centering
	\includegraphics[keepaspectratio=true,width=0.48\textwidth]{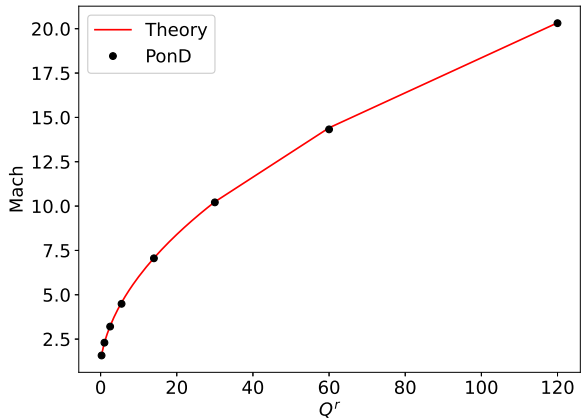}
	\caption{{Mach number v/s heat of reaction for 1D detonation. $Q^{\rm r}$ is the heat of reaction.} } 
	\label{fig:1d_detonation_mach}
\end{figure}

\begin{figure}[ht]
	\centering
	\includegraphics[keepaspectratio=true,width=0.48\textwidth]{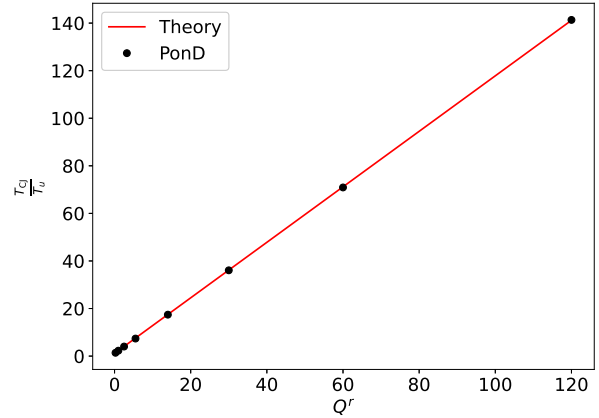}
	\caption{Temperature at the Chapman–-Jouguet point v/s heat of reaction for 1D detonation. $Q^{\rm r}$ is the heat of reaction.} 
	\label{fig:1d_detonation_temperature}
\end{figure}

\begin{figure}[ht]
	\centering
	\includegraphics[keepaspectratio=true,width=0.48\textwidth]{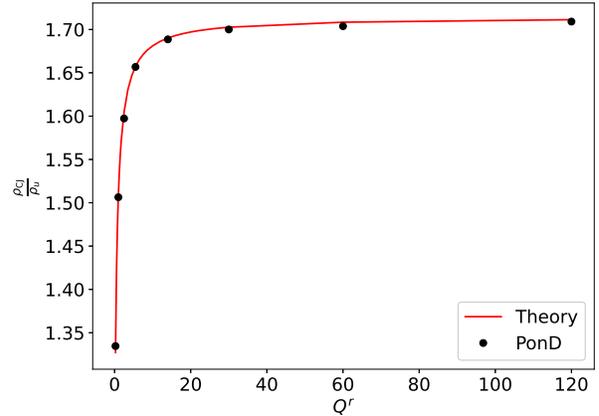}
	\caption{Density at the Chapman–-Jouguet point v/s heat of reaction for 1D detonation. $Q^{\rm r}$ is the heat of reaction.} 
	\label{fig:1d_detonation_density}
\end{figure}

The Mach numbers of the detonation waves obtained from our simulations at different $Q^{\rm r}$ are presented in Figure (\ref{fig:1d_detonation_mach}). At the CJ point corresponding to those Mach numbers, the measurements  of temperature ratios $T_{\rm CJ}/T_u$ and density ratios $\rho_{\rm CJ}/\rho_u$ have been shown in Figure (\ref{fig:1d_detonation_temperature}) and Figure (\ref{fig:1d_detonation_density}), respectively. The profiles of density and temperature for $Q^{\rm r}=14$ are shown in Figure (\ref{fig:1d_detonation_profiles}). The results show that the proposed method has not only proved to be stable over a wide range of Mach numbers and temperature ratios but it has also captured the detonation hydrodynamics with excellent accuracy. Thus, the Particle on Demand method significantly extends the applicability of the traditional lattice Boltzmann method to higher Mach numbers and larger temperature ratios. Having obtained good quantitative agreement in the simplest one-dimensional setup, we proceed to a detonation experiment in two-dimensions to verify the accuracy in two dimensions as well as to check the isotropy of the proposed method.
\subsection{\label{sec:2d_circular_detonation}2D Circular detonation}
\begin{figure}[ht]
	\centering
	\includegraphics[keepaspectratio=true,width=0.48\textwidth]{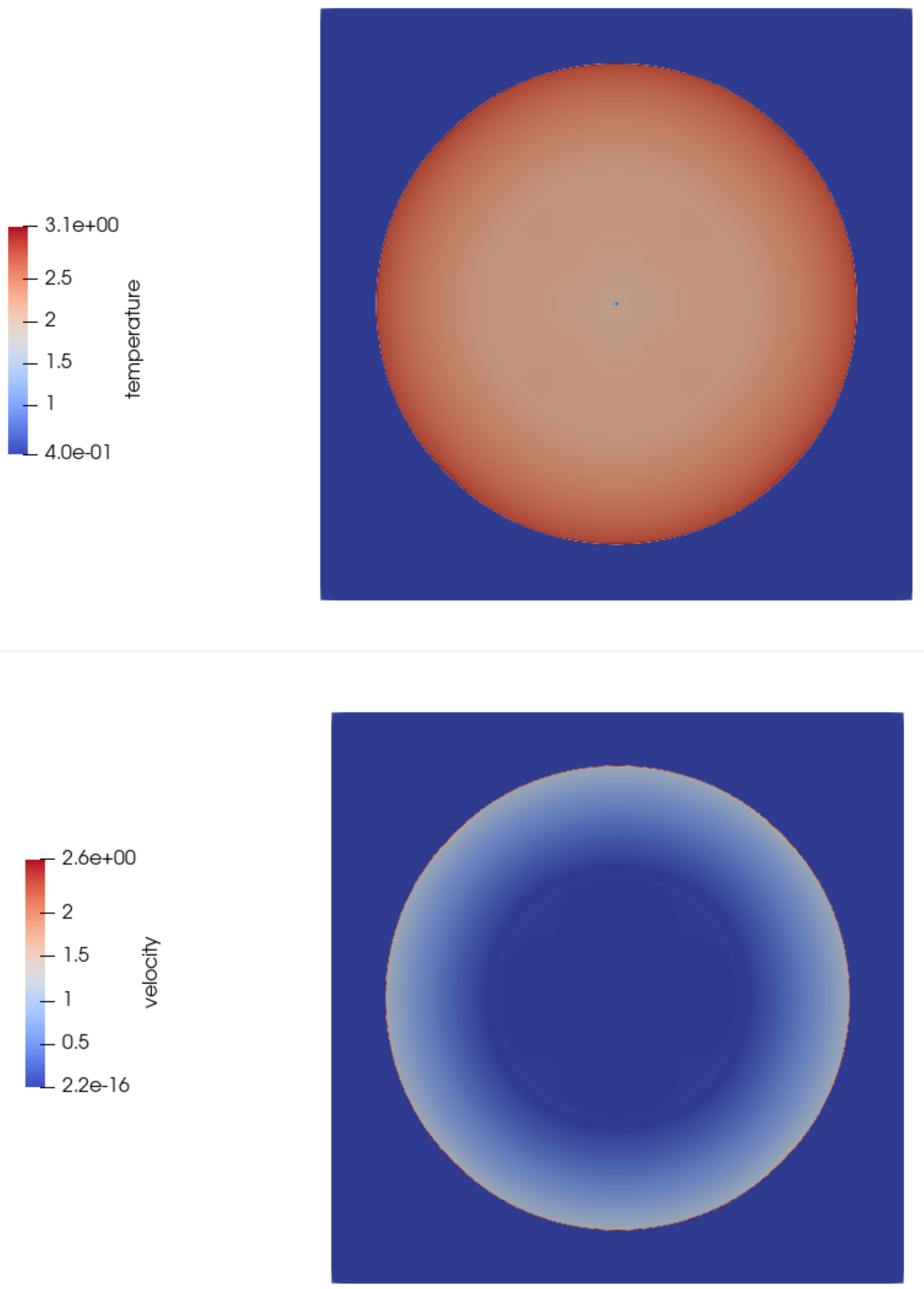}
	\caption{Temperature and velocity for 2D circular detonation.} 
	\label{fig:2d_circular_detonation}
\end{figure}

\begin{figure}[ht]
	\centering
	\includegraphics[keepaspectratio=true,width=0.49\textwidth]{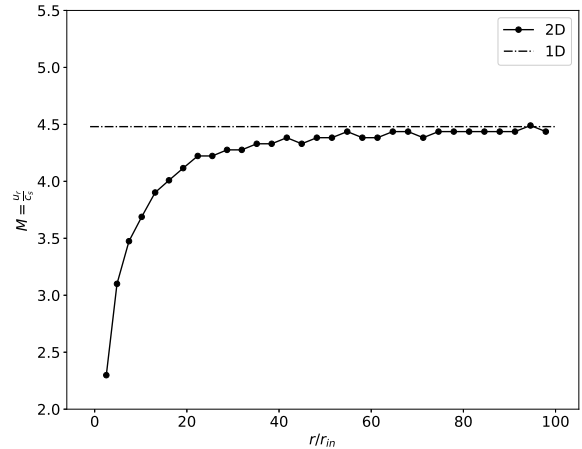}
	\caption{Mach number of the detonation front v/s radius. The horizontal dotted line represents $M=4.48$ achieved at the steady state in one-dimensional simulation. The continuous lines show the wave speed as a function of the location of the detonation wavefront in the two-dimensional circular detonation simulation.} 
	\label{fig:2d_circular_detonation_mach}
\end{figure}

An outward expanding circular detonation in a square $L \times L$ area is computed  with $N_x \times N_y = 5000 \times 5000$ lattice nodes. The centre of the computational domain is initialized with a circular hot spot of radius $r_{in}=L/200$. The hot spot is initialized with burnt conditions corresponding to the product mass fraction $\Lambda_b$, pressure $P_b$ and density $\rho_b$. Except for the hot spot, the remaining computational domain is initialized with unburnt conditions given by product mass fraction $\Lambda_u$ at a pressure $P_u$ and density $\rho_u$. 
\begin{align}
(\Lambda,\rho,P)=\begin{cases}
\Lambda_u=0,\rho_u=1.00,P_u=0.4,&r > r_{in},\\
\Lambda_b=1,\rho_b=1.65,P_b=1.0,&\text{otherwise.}
\label{eq:ic2dDet}
\end{cases}
\end{align}
We compute this setup at the non-dimensional heat of reaction $Q^{\rm r}=5.50$.
In an outward expanding circular or spherical detonation, the radial speed of the detonation wavefront is expected to approach the analytical one dimensional detonation wave speed as the radius of the expanding wavefront approaches infinity \cite{taylor_dynamics_1950}. The radial Mach number of the expanding wavefront from the simulation is plotted against the radius of the wavefront in Figure (\ref{fig:2d_circular_detonation_mach}). In the figure, the detonation wave is seen to be accelerating with increasing radius while finally  approaching the one dimensional detonation wave speed $M=4.48$. Figure (\ref{fig:2d_circular_detonation}) shows the contours of temperature and magnitude of the velocity at a time instant when the solution has almost achieved the one dimensional detonation speed. As the solution is free of visible artefacts from the underlying Cartesian grid or the lattice, it can be inferred that the model does not suffer from problems of unphysical anisotropy. In the next example, we perform a stricter quantitative check of isotropy by measuring the critical angle of transition to Mach stems between opposed interacting detonation wavefronts.
\subsection{\label{sec:2d_mach_reflection}2D Mach reflection}
In the event of an incident shock wave producing a reflected shock from a solid boundary, there exists a critical angle $\phi_c$ between the incident shock and
the solid boundary, above which a fluid parcel cannot pass through both the incident and the reflected shocks while maintaining its original trajectory \cite{yirak_mach_2013}. Once the critical angle is exceeded, the solution of the flow evolves into a new configuration consisting of an additional third shock which is normal to the solid boundary. The new shock, known as the Mach stem, pushes the point of the intersection of the incident and the reflected shock away from the solid surface, thereby forming a triple point. 
In case of a strong shock, the critical angle  can be approximated as a function of the adiabatic constant  $\gamma$ by the relation \cite{courant_supersonic_1999},
\begin{align}
    \phi_c = \arcsin\left( \frac{1}{\gamma} \right).
    \label{eq:critical_angle}
\end{align}
In the absence of a solid body, a Mach stem can be produced by the intersection of two co-moving bow shocks \cite{yirak_mach_2013}. In such a configuration, the line of symmetry traced by the motion of the intersecting point between the bow shocks assumes the role of a solid boundary. Here, we produce Mach stems using the interaction of two strong detonation shocks. The computational domain is made up of $N_x \times N_y = 600 \times 400$ lattice nodes representing a rectangular area $L_x \times L_y$. The area is initialized with unburnt conditions corresponding to the product mass fraction $\Lambda_u=0$ at a pressure and density of $P_u=0.4$ and $\rho_u=1$, respectively. The left top and left bottom corners are initialized with quarter circular hot spots of radius $r_{in}=L_x/20$. The hot spots are initialized with burnt conditions corresponding to the product mass fraction $\Lambda_b=1$ at a pressure and density of $P_b=1.0$ and $\rho_b=1.65$, respectively. 
The non-dimensional heat of reaction is $Q^{\rm r}=30.0$. A large value of $Q^{\rm r}$ has been selected in order to create strong shocks of high pressure ratios. With an intent to verify the critical angles predicted by (\ref{eq:critical_angle}), we compute this setup for two different values of adiabatic exponents, $\gamma=1.4$ and $\gamma=5/3$. 

\begin{figure}[ht]
	\centering
	\includegraphics[keepaspectratio=true,width=0.48\textwidth]{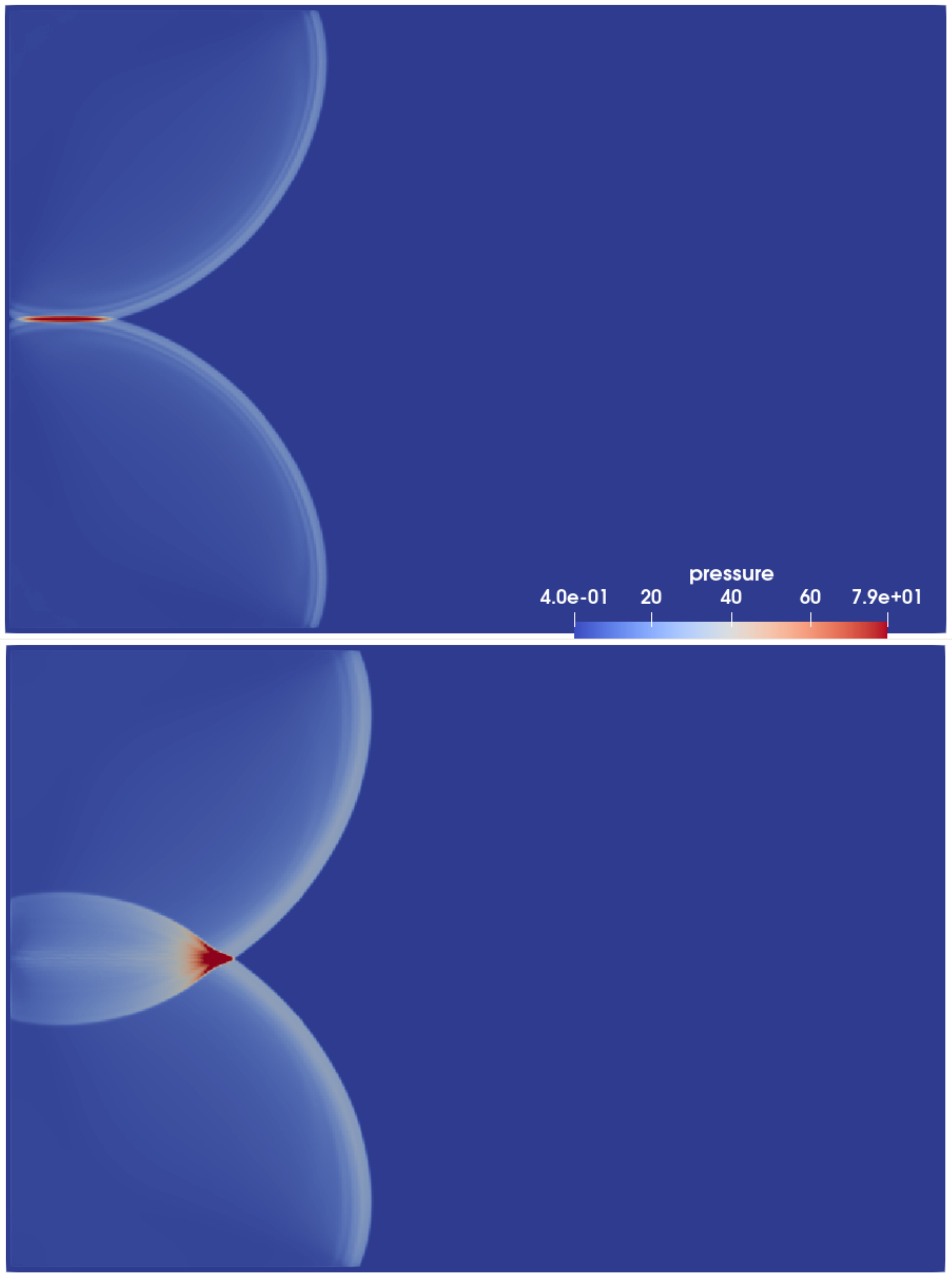}
	\caption{
	Incident detonation waves forming regular reflection waves. PonD simulations with adiabatic exponents $\gamma=1.4$ (top) and $\gamma=5/3$ (bottom). t=44.} 
	\label{fig:2d_mach_reflection_t44}
\end{figure}

\begin{figure}[ht]
	\centering
	\includegraphics[keepaspectratio=true,width=0.48\textwidth]{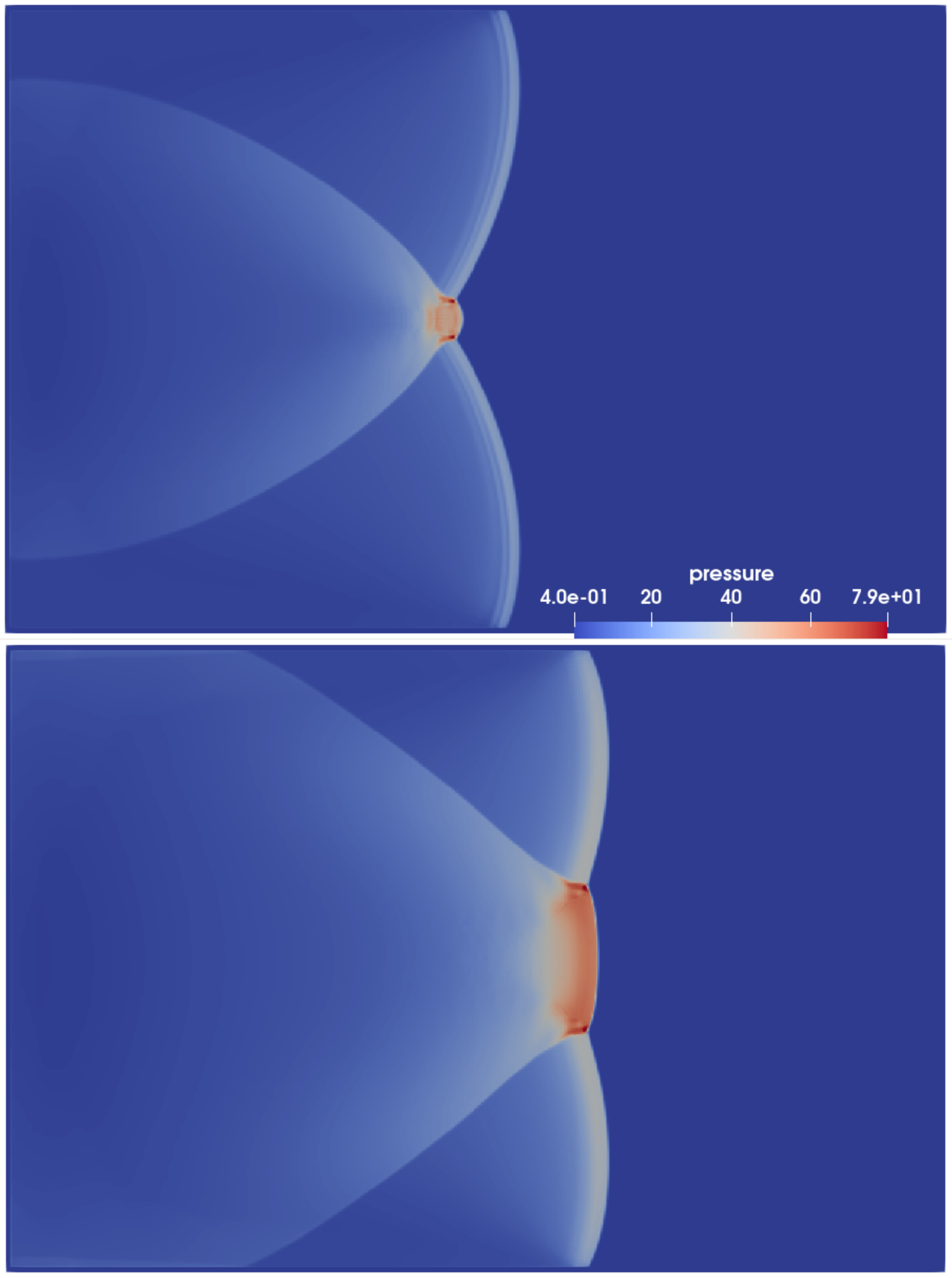}
	\caption{
	Mach reflection and regular reflection created from the interaction of incident detonation waves. PonD simulations with adiabatic exponents $\gamma=1.4$ (top) and $\gamma=5/3$ (bottom). t=68.
    } 
	\label{fig:2d_mach_reflection_t68}
\end{figure}

From the initial hot spots, detonation waves propagate outward towards each other, intersect and then continue propagating. In the top frame of Figure (\ref{fig:2d_mach_reflection_t44}), the solution for $\gamma=1.4$ shows meeting of the opposite detonation waves. At the same time instant, the faster ``incident" detonation waves for $\gamma=5/3$ have already formed reflected waves know as the waves of ``regular reflection". The angle between the incident waves and the horizontal symmetry line is still smaller than the critical angle $\phi_c$ and therefore the solution at $t=44$ consists of only incident shock and the regular reflected shock. Due to the ongoing motion of the incident bow shocks, the included angle continues to increase until the critical angle is reached. After exceeding the critical angle, a Mach stem is formed as expected, resulting into a three-shock structure as is visible in Figure (\ref{fig:2d_mach_reflection_t68}) for the $\gamma=1.4$ as well as the $\gamma=5/3$ simulation. Good agreement with the analytical relation (\ref{eq:critical_angle}) is obtained for the critical angles in both simulations, as reported in Table (\ref{tab:machReflection}).

Having obtained good quantitative agreement in one and two dimensions, we present a three dimensional simulation to show that present model is also realizable and feasible for three dimensional simulations in principle.
\begin{table}[h!]
\centering
\begin{tabular}{ |p{1.5cm}||p{1.7cm}|p{1.5cm}|  }
 \hline
 \multicolumn{3}{|c|}{ Mach reflection } \\
 \hline
 $\gamma$ & $\arcsin\left({1}/{\gamma} \right)$ & $\phi_c$ PonD \\
 \hline
 $1.4$   &   $45.58$  & $45.48$\\
 $5/3$   &   $36.86$  & $36.20$\\
 \hline
\end{tabular}
\caption{Critical angles v/s $\gamma$ for Mach reflection.}
\label{tab:machReflection}
\end{table}
\subsection{\label{sec:3d_spherical_detonation}Spherical detonation}
\begin{figure}[h!]
	\centering
	\includegraphics[trim={0cm 9cm 0cm 9cm},clip,keepaspectratio=true,width=0.48\textwidth]{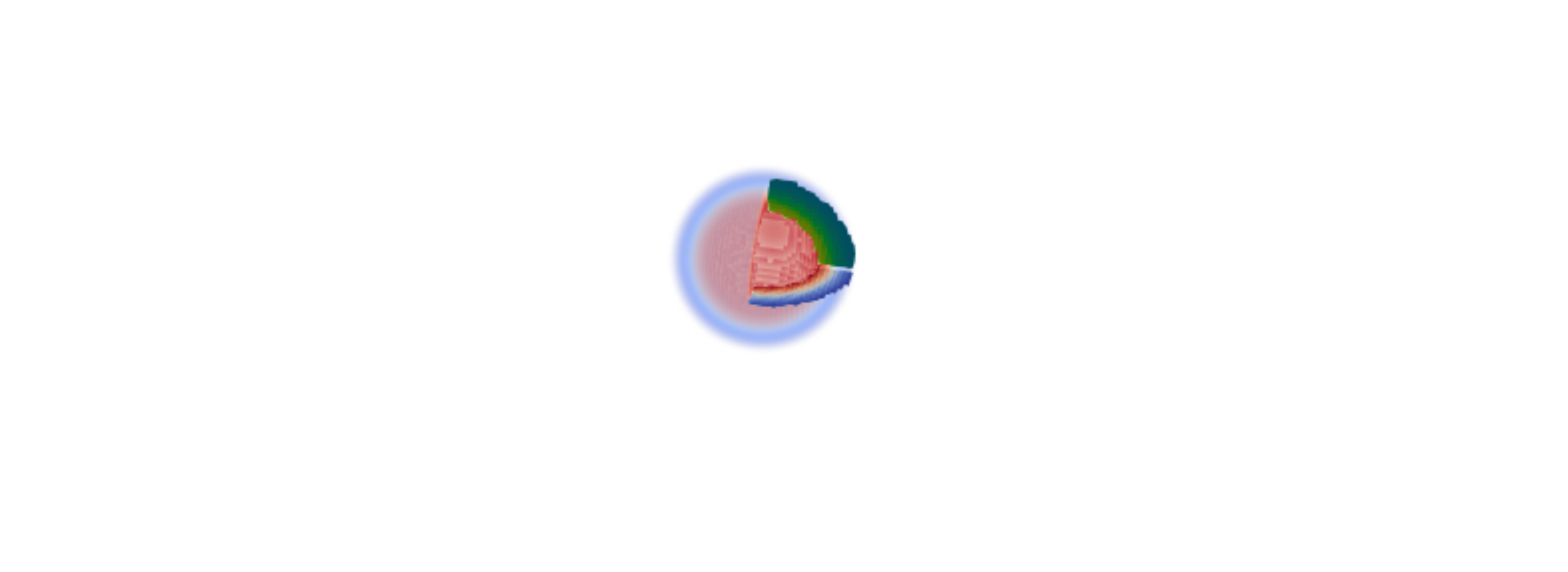}
	%\hline
%trim={<left> <lower> <right> <upper>}
	\includegraphics[trim={0cm 4cm 0cm 4cm},keepaspectratio=true,width=0.48\textwidth]{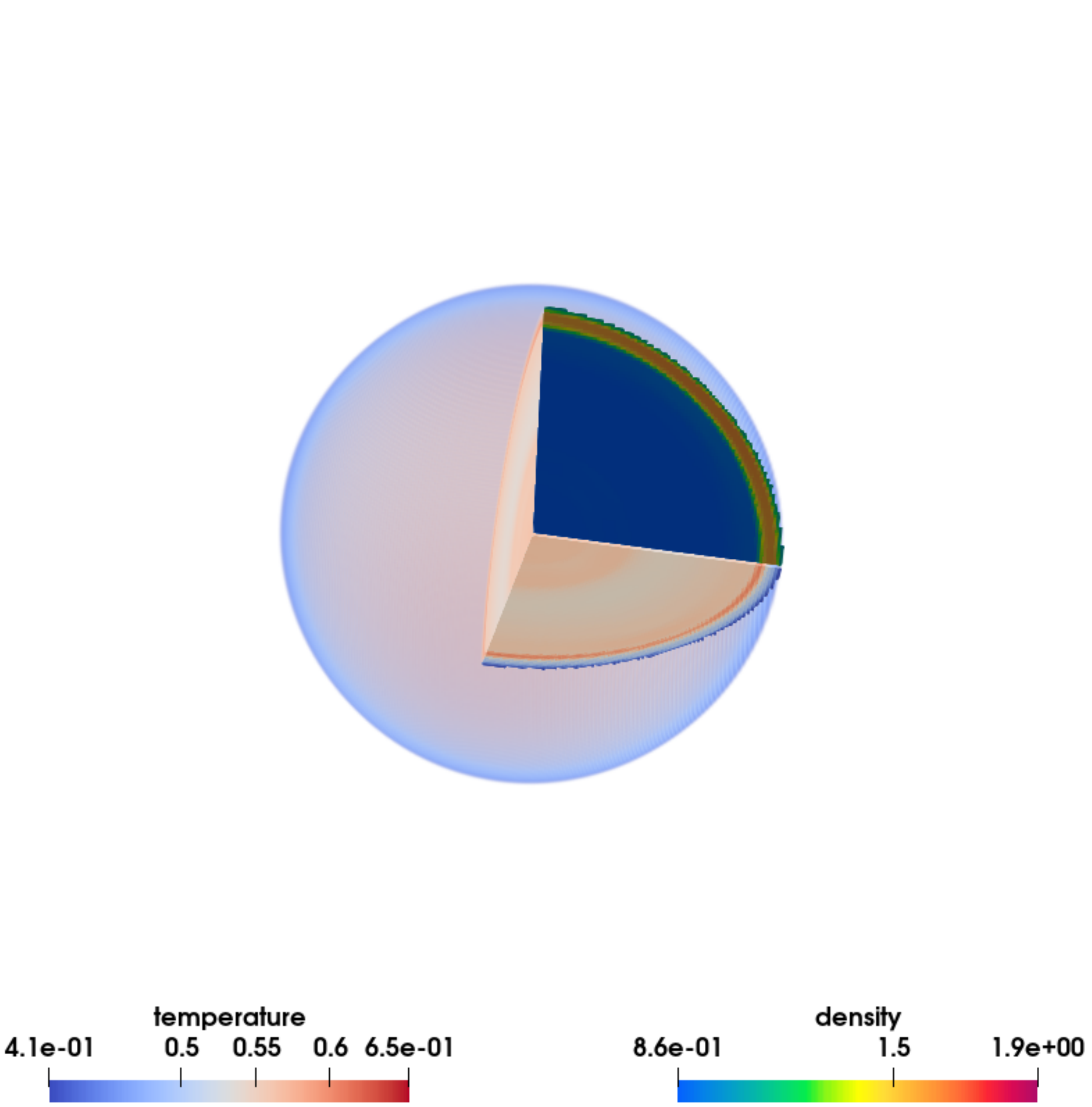}
	%\hline
	\includegraphics[keepaspectratio=true,width=0.48\textwidth]{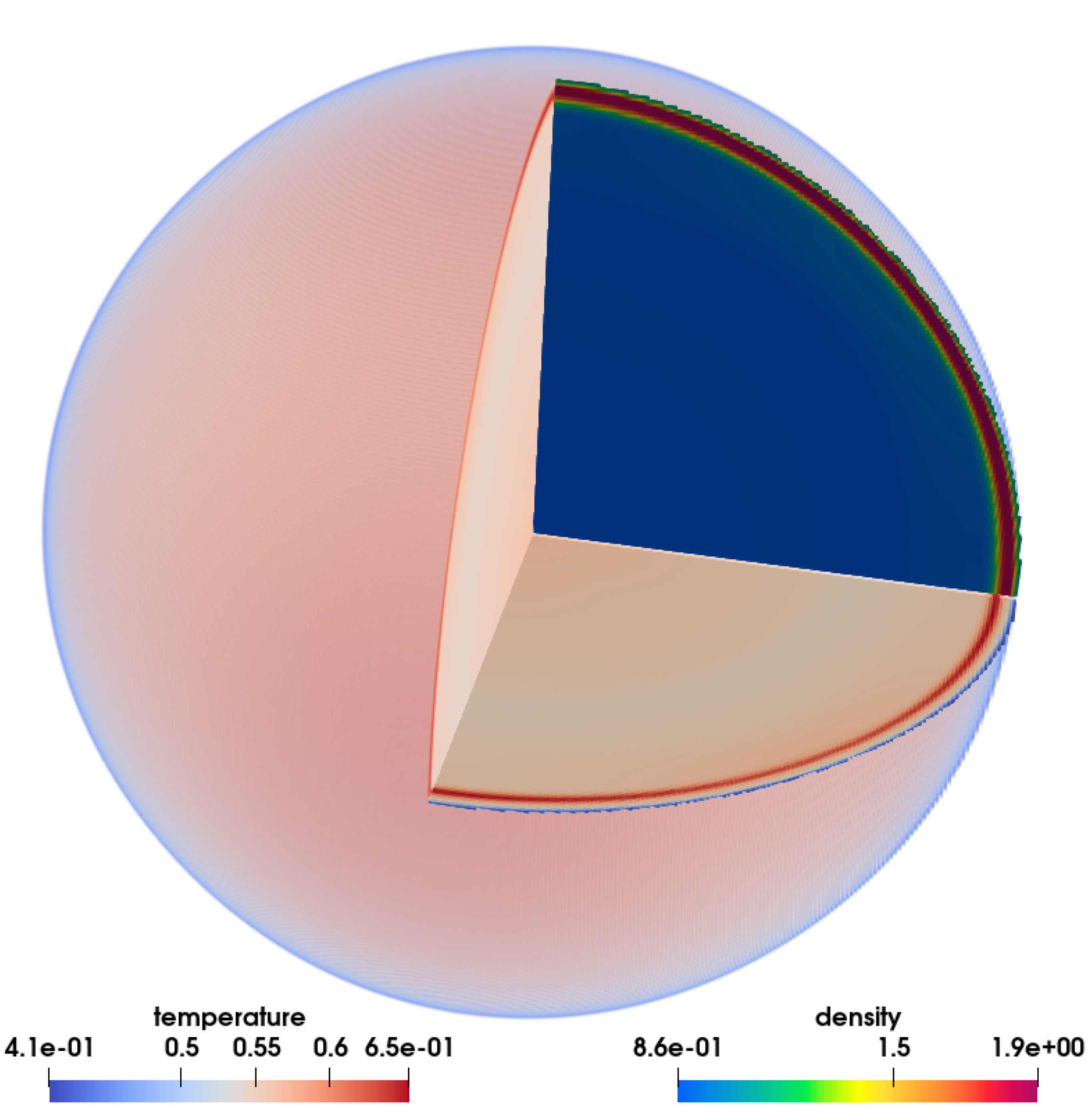}
	\caption{Temperature iso-volume and	density slice for a  spherical detonation at different times. From top to bottom: $t=0$, $t=8$, $t=16$. } 
	\label{fig:3d_spherical_detonation_t16}
\end{figure}
\begin{figure}
	\centering
	\includegraphics[keepaspectratio=true,width=0.48\textwidth]{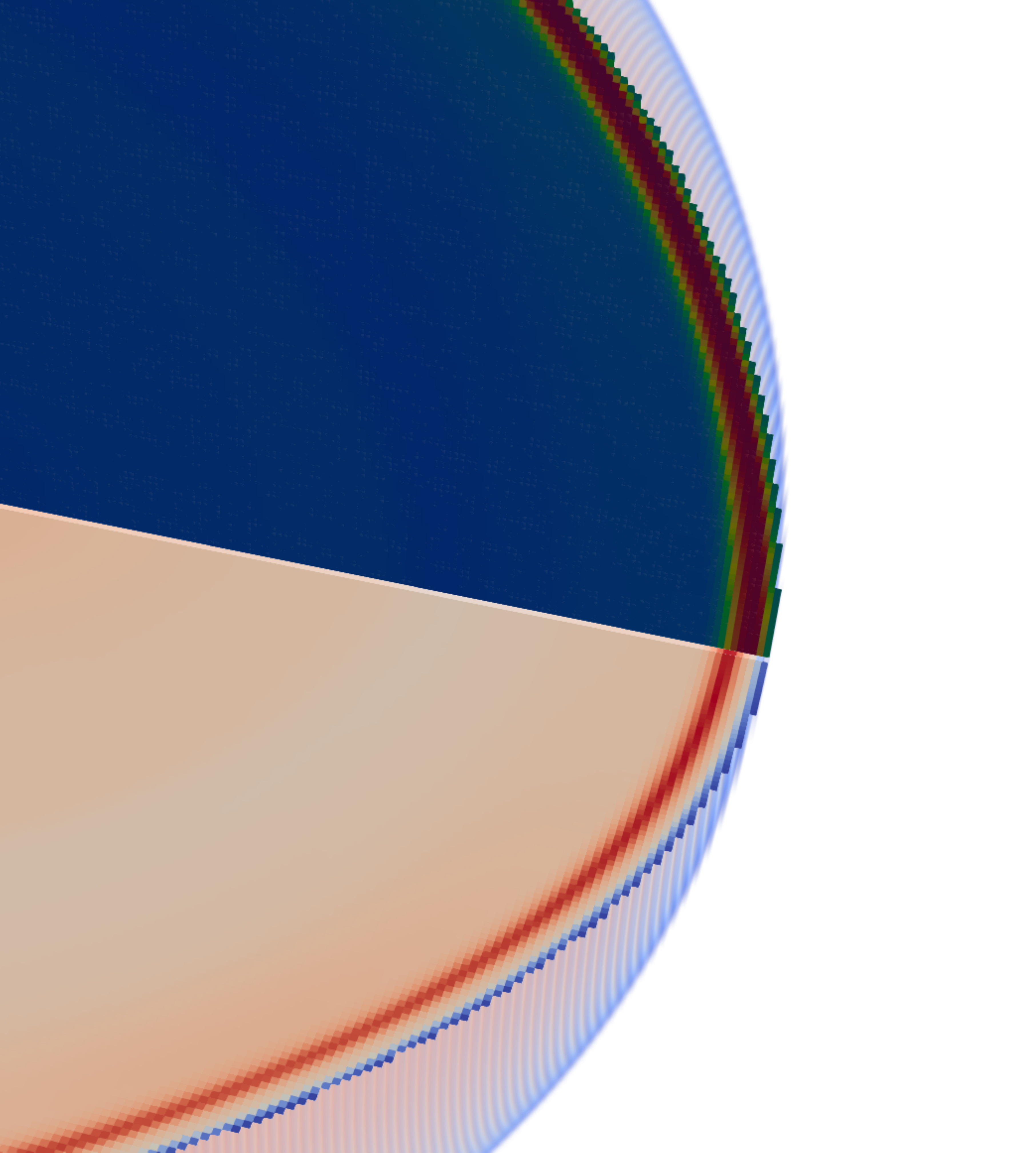}
	\caption{Close up of the temperature slice (horizontal) and the density slice (vertical) in  spherical detonation. $t=16$. The contour levels and colors are unchanged from Figure \ref{fig:3d_spherical_detonation_t16}.} 
	\label{fig:3d_spherical_znd}
\end{figure}
A spherical detonation simulation is performed with an implementation of the proposed model on the three dimensional $D3Q39$ lattice \cite{frapolli_theory_2020}. As evident from two-dimensional simulations of the circular expanding detonation in section \ref{sec:2d_circular_detonation}, a very large domain is necessary for the multi-dimensional solution to approach the one-dimensional solution. In order to avoid the prohibitive computational cost associated with a large three dimensional simulation, we perform only a qualitative simulation in a small cubic domain $L \times L \times L$ consisting of $N_x \times N_y \times N_z = 400 \times 400 \times 400$ lattice nodes. 

The computational domain is initialized with unburnt conditions corresponding to the product mass fraction $\Lambda_u$ at a pressure and density of $P_u$ and $\rho_u$, respectively. The centre of the domain is initialized with a spherical hot spot of diameter $d_{in}=0.075 L$. We simulate this setup for the non-dimensional heat of reaction $Q^{\rm r}=0.50$. The product mass fraction in the hot spot is initialized with the burnt state $\Lambda_b$, while the pressure $P_b$ and density $\rho_b$ are set to the analytical CJ state values corresponding to $Q^{\rm r}$, 
\begin{align}
(\Lambda,\rho,P)=\begin{cases}
\Lambda_u=0,\rho_u=1.00,P_u=0.40,&d > d_{in},\\
\Lambda_b=1,\rho_b=1.41,P_b=0.96,&\text{otherwise.}
\label{eq:ic3dDet} 
\end{cases}
\end{align}
The kinematic viscosity is $\nu=10^{-5}$. The initial condition is shown in Figure (\ref{fig:3d_spherical_detonation_t16}) (top) by an iso-volume of temperature. A quarter of the volume has been clipped away to show the temperature and the density distribution inside the spherical volume. The subsequent evolution of the solution is also shown in Figure (\ref{fig:3d_spherical_detonation_t16}). As is evident from the results, just as in the two dimensional simulations, the proposed method maintains the roundness of the initial conditions also in three dimensions. As expected, the underlying Cartesian grid or the lattice does not produce any unphysical imprint on the solution. In the  Zel'dovich, von Neumann and Döring (ZND) wave structure model \cite{williams_combustion_1985} for detonation, a shock wave heats the reactants to a temperature high enough to initiate reaction. This results into an exothermic reaction occurring immediately behind the shock, which in turn provides energy to sustain the shock. The vertical slice in Figure (\ref{fig:3d_spherical_detonation_t16}) reveals that a layer (or shell) of high density i.e. a hydrodynamic shock forms the outermost layer of the expanding sphere. The horizontal slice reveals the circumference of the shell of maximum temperature which is associated with the exothermic chemical reaction. A closeup of the intersection of the density and the temperature slice in Figure (\ref{fig:3d_spherical_znd}) shows that the shell of maximum temperature immediately follows the shell of maximum density. Thus, the computation evolved into spherical shells that form ZND structures but in three dimensions. 
%At the time instant $t=16$, the detonation wave attained a Mach number of $1.4$ and expanded to a size $d=9.5 d_{in}$. At that time instant, the maximum temperature in the domain was $T=1.625 T_u$ and the maximum density was $\rho=1.9 \rho_u$. The size of the domain was insufficient for the sphere to reach the theoretical Mach number of $1.84$ predicted from the CJ theory.      
%
%
\section{\label{sec:conclusion}Conclusion}
The Particle on Demand (PonD) method was combined with the Lee--Tarver detonation model to show the applicability of the Pond method to detonation regime hydrodynamics. Simulations with the combined model show good agreement with the theoretical density and temperature predictions for one dimensional detonation waves as fast as Mach 20. Two dimensional circular detonation computation is not only free of unphysical anisotropy but also produces the correct propagation speed for the detonation wavefront. Critical angles of Mach stems are also validated in two dimensions.  A spherical detonation simulation is performed as a feasibility check for setups of practical interest. The proposed model is a fitting prequel to future realistic supersonic combustion modelling using PonD for hydrodynamics with detailed chemistry models \cite{sawant_consistent_2021,sawant_lattice_2021,sawant_consistent_2021-1}.
\begin{acknowledgments}
The authors thank Nikolaos Kallikounis for discussions on PonD in general and the Grad's projection based transformation in specific. 
This work was supported by European Research Council (ERC) Advanced Grant no. 834763-PonD  Computational resources at the Swiss National Super Computing Center CSCS were provided under grant no. s1066.
\end{acknowledgments}
\appendix

\section{Moments of the distribution functions}
\label{sec:moments:appendix}
The moments of the distribution functions $f_i$ and $g_i$ are,
\begin{align}
    &\sum_i f_i^{\rm eq} = \sum_i f_i = \rho \\
    &\sum_i f_i^{\rm eq} v_{i\alpha} = \sum_i f_i v_{i\alpha} = \rho u_\alpha\\
    &\sum_i f_i^{\rm eq} v_{i\alpha} v_{i\beta}= \rho (u_\alpha u_\beta + T \delta_{\alpha \beta}) \\
    &\sum_i f_i^{\rm eq} v_{i\alpha} v_{i\beta} v_{i\gamma}= \rho u_\alpha u_\beta u_\gamma + \rho T (u_\alpha \delta_{\beta \gamma} + u_\beta \delta_{\gamma \alpha} + u_\gamma \delta_{\alpha \beta}) \\
% \end{align}
% %
% \begin{align}
    &\sum_i g_i^{\rm eq} + \sum_i f_i^{\rm eq} v_{i\gamma} v_{i\gamma} =  \\
    &\sum_i g_i + \sum_i f_i v_{i\gamma} v_{i\gamma} = 2 \rho \left(C_v T + \frac{u_\gamma u_\gamma}{2}\right)=2 \rho E \nonumber \\
    &\sum_i g_i^{\rm eq} v_{i\alpha} + \sum_i f_i^{\rm eq} v_{i\gamma} v_{i\gamma} v_{i\alpha} = 2 \rho u_\alpha (E+T) \\
    &\sum_i g_i^{\rm eq} v_{i\alpha} v_{i\beta} + \sum_i f_i^{\rm eq} v_{i\gamma} v_{i\gamma} v_{i\alpha} v_{i\beta}= 2 \rho T u_\alpha u_\beta \nonumber \\
    & + 2 \rho (u_\alpha u_\beta + T \delta_{\alpha \beta}) (E+T)
\end{align}
\section{Coefficients of the Hermite polynomial}
\label{sec:hermiteCoefficents:appendix}
For moments up to the third order in the particle velocity space,
\begin{align}
    \sum_i f_i &= M^{(0)}, \label{eq:genericMomentO0}\\
    \sum_i f_i v_{i\alpha}(u_\alpha,T) &= M_\alpha^{(1)}, \\
    \sum_i f_i v_{i\alpha}(u_\alpha,T) v_{i\beta}(u_\beta,T) &= M^{(2)}_{\alpha\beta}, \\
    \sum_i f_i v_{i\alpha}(u_\alpha,T) v_{i\beta}(u_\beta,T) v_{i\gamma}(u_\gamma,T) &= M^{(3)}_{\alpha\beta\gamma},
    \label{eq:genericMomentO3}
\end{align}
The expansion is given by the series,
\begin{align}
    &f_i = f_i^{(0)} \sum_{n=0}^{\infty} \frac{1}{n!} a(\bm{m};\lambda(\bm{u},T))^{(n)} H_i^{(n)}
    \label{eq:gradGeneralAppendix}
\end{align}
The polynomials of the series upto the third order are, 
\begin{align}
    &H_i^{(0)} = 1 \\
    &H_{i\alpha}^{(1)} = c_{i\alpha} \\
    &H_{i\alpha \beta}^{(2)} = c_{i\alpha} c_{i\beta} - T_L \delta_{\alpha \beta} \\
    &H_{i\alpha \beta \gamma}^{(3)} = c_{i\alpha} c_{i\beta} c_{i\gamma} - T_L (c_{i\alpha} \delta_{\beta \gamma} + c_{i\beta} \delta_{\gamma \alpha} + c_{i\gamma} \delta_{\alpha \beta}) 
\end{align}
The corresponding coefficients are,
\begin{align}
    &a^{(0)}= M^{(0)} \\
    &a_\alpha^{(1)} = \frac{1}{\tilde{T} T_L}\left( M_\alpha^{(1)} - u_\alpha M^{(0)} \right) \\
    &a_{\alpha\beta}^{(2)} = \frac{1}{\tilde{T}^2 T_L^2}[ M_{\alpha \beta}^{(2)} - M^{(0)} \tilde{T}^2 T_L \delta_{\alpha \beta} - M^{(0)} u_\alpha u_\beta \\ \nonumber
    &- u_\alpha ( M_\beta^{(1)} - u_\beta M^{(0)} )  - u_\beta ( M_\alpha^{(1)} - u_\alpha M^{(0)}) ] \\
    &a_{\alpha\beta\gamma}^{(3)}=\frac{1}{\tilde{T}^3 T_L^3}[ M^{(3)}_{\alpha\beta\gamma} - M^{(0)} u_\alpha u_\beta u_\gamma \\ \nonumber &-(M_\alpha^{(1)}-M^{(0)}u_\alpha)(T_L \tilde{T}^2 \delta_{\beta\gamma} - u_\beta u_\gamma) \\ \nonumber
    &-(M_\beta^{(1)}-M^{(0)}u_\beta)(T_L \tilde{T}^2 \delta_{\gamma\alpha} - u_\gamma u_\alpha) \\ \nonumber
    &-(M_\gamma^{(1)}-M^{(0)}u_\gamma)(T_L \tilde{T}^2 \delta_{\alpha\beta} - u_\alpha u_\beta) \\ \nonumber
    &-(M^{(2)}_{\alpha\beta}-M^{(0)}u_\alpha u_\beta)u_\gamma \\ \nonumber
    &-(M^{(2)}_{\beta\gamma}-M^{(0)}u_\beta u_\gamma)u_\alpha \\ \nonumber
    &-(M^{(2)}_{\gamma\alpha}-M^{(0)}u_\gamma u_\alpha)u_\beta 
    ]
\end{align}
\section{Analytical solutions}
\label{sec:analytical_solution:appendix}
The Rankine-Hugoniot equations relate the properties on the upstream and downstream sides of combustion waves in infinite, plane, steady-state, one-dimensional flows involving exothermic chemical reactions \cite{williams_combustion_1985,law_combustion_2006}. Let us denote the conditions upstream of the shock by subscript (0) and the post-combustion downstream conditions by subscript (1).
Using the Rankine-Hugoniot equations with the non-dimensional heat of reaction $\alpha$ defined as,
\begin{align}
    &\alpha = Q^{\rm r} \frac{\rho_0}{P_0}
\end{align}
The Mach number of the detonation wavefront is estimated as a function of heat of reaction and the adiabatic exponent $\gamma$,
\begin{align}
    M_{1} = \sqrt{1+\frac{\alpha (\gamma^2-1)}{2 \gamma}} + \sqrt{\frac{\alpha (\gamma^2-1)}{2 \gamma}} 
\end{align}
Using the shock velocity $u_1=M_1 \sqrt{\gamma T_0}$, the relative downstream (post-shock) velocity $u_2$ is given by,
\begin{align}
    &u_2 = u_1 - u_{\rm local}
\end{align}
The downstream location at which the relative velocity is equal to the local speed of sound is called the Chapman-Jouguet (CJ) point. At the location of the CJ point $x_{\rm CJ}$,
\begin{align}
    &u_2(x_{\rm CJ}) = \sqrt{\gamma T (x_{\rm CJ})}
\end{align}
The pressure and density ratios at $x_{\rm CJ}$ predicted by the Rankine-Hugoniot equations are, respectively,
\begin{align}
    &\frac{P_{1}}{P_0} = 1 + \alpha (\gamma - 1) \left(1 + \sqrt{1+\frac{2 \gamma}{\alpha (\gamma^2 -1 )}} \right)
    \\
    &\frac{\rho_0}{\rho_{1}} = 1 + \alpha \left(\frac{\gamma - 1}{\gamma}\right) \left(1 - \sqrt{1+\frac{2 \gamma}{\alpha (\gamma^2 -1 )}} \right) 
\end{align}
\section{Lattice specifications}
\label{sec:lattices:appendix}
\begin{table}[h!]
\centering
\begin{tabular}{ |p{2.5cm}||p{2cm}|p{1.2cm}|  }
 \hline
 \multicolumn{3}{|c|}{ D2Q16, $T_L=1$ } \\
 \hline
 $(c_{ix},c_{iy}) $ & $w_i$ & $i$ \\
 \hline
 $(\pm1,\pm1)$   &   $\frac{3+\sqrt{6}}{12} \times \frac{3+\sqrt{6}}{12}$  & $0-3$\\
 $(\pm1,\pm2),(\pm2,\pm1)$   &   $\frac{3+\sqrt{6}}{12} \times \frac{3-\sqrt{6}}{12}$  & $4-11$\\
 $(\pm2,\pm2)$   &   $\frac{3-\sqrt{6}}{12} \times \frac{3-\sqrt{6}}{12}$  & $12-15$\\
 \hline
\end{tabular}
\caption{The D2Q16 Lattice.}
\label{tab:theD2Q16Lattice}
\end{table}
\begin{table}[h!]
\centering
\begin{tabular}{ |p{4.8cm}||p{1.05cm}|p{1.0cm}|  }
 \hline
 \multicolumn{3}{|c|}{ D3Q39, $T_L=2/3$ } \\
 \hline
 $(c_{ix},c_{iy},c_{iz}) $ & $w_i$ & $i$ \\
 \hline
 $(0,0,0)$   &   $1/12$  & $0$\\
 $(\pm1,0,0),(0,\pm1,0),(0,0,\pm1)$   &   $1/12$  & $1-6$\\
 $(\pm2,0,0),(0,\pm2,0),(0,0,\pm2)$   &   $2/135$  & $7-12$\\
 $(\pm3,0,0),(0,\pm3,0),(0,0,\pm3)$   &   $1/1620$  & $13-18$\\
 $(\pm2,\pm2,0),(0,\pm2,\pm2),(\pm2,0,\pm2)$   &   $1/432$  & $19-30$\\
 $(\pm1,\pm1,\pm1)$   &   $1/27$  & $31-38$\\
 \hline
\end{tabular}
\caption{The D3Q39 Lattice.}
\label{tab:theD3Q39Lattice}
\end{table}

%\nocite{*}
\bibliography{library}% Produces the bibliography via BibTeX.

\end{document}